# Measurement of energy gaps in superconductors by means of quantum interference devices

S. I. Bondarenko, V. P. Koverya, A. V. Krevsun, and L. V. Gnezdilova

*B. I. Verkin Institute of Low-Temperature Physics and Engineering, National Academy of Sciences of Ukraine, pr. Lenina 47, Kharkov 61103, Ukraine*

The effect of temperature on the form of discrete changes in the current in highly inductive ($\sim 10^{-6}$ H) doubly-connected superconductors with niobium-niobium clamped point contacts is determined experimentally. The magnitude and duration of the voltage pulse on a doubly-connected superconductor is measured at the time of the discrete change in its current state. The pulse magnitude is close to the energy gap $2\Delta/e$ of the superconductor and its duration ($\sim 10^{-6}$ s) corresponds to the minimum possible time ($\sim 10^{-12}$ s) for a change in the state of the contact when the depairing current through it is reached. The measurement data are discussed in terms of models of the quantum interference of currents in a doubly-connected superconductor with clamped point contacts in the form of a quantum interferometer.

## 1. Introduction

Periodic, as well as discrete, changes in low-induction ($L_0 < 10^{-9}$ H at $T = 4.2$ K) superconducting circuits in the form of a ring with point contacts[1-3] in an external magnetic field were discovered in the 1960s. The reason for these changes was quantization of the z-component of the canonical momentum ($M_z$) of Cooper pairs moving about the Z axis along a closed superconducting circuit of radius $r$:

$$\oint M_z d\varphi = \oint \mathbf{p} r d\varphi = \oint \mathbf{p} d\mathbf{l} = Nh, \qquad (1)$$

where $\mathbf{p}$, $h$, and $N$ are, respectively, the pair momentum, Planck constant, and an integer. Given that $\mathbf{p} = 2m\mathbf{v} + 2e\mathbf{A}$ (where $2m, \mathbf{v}, 2e$ are the mass, velocity, and charge of the Cooper pair and $\mathbf{A}$ is the vector potential of the magnetic field) in a magnetic field perpendicular to the plane of the ring, Eq. (1) can be used to obtain the relationship between the circulating current $i$ produced in this circuit and the magnetic flux $\Phi$ through the circuit.[2] In a linear approximation this relationship has the form

$$i = (n\Phi_0 - \Phi)/\gamma L_0, \qquad (2)$$

where $\gamma = (m/ne^2)(t/\sigma L_0)$, $\Phi_0$ is the quantum of magnetic flux $h/2e$, $n$ is the density of Cooper pairs, $t$ and $\sigma$ are the length and cross section of the point contact in the circuit, and $L_0$ is the inductance of the circuit. In the nonlinear Josephson approximation this relationship is given by

$$i = -i_c \sin[2\pi(L_0 i + \Phi_x - n\Phi_0)/\Phi_0], \qquad (3)$$

where $\Phi_x = \Phi - L_0 i$ and $\Phi_x$ is the magnetic flux of the external magnetic field. In the linear approximation, as opposed to the Josephson approximation, the density of Cooper pairs ($n$) is independent of the current. In the linear approximation Eq. (2) implies that for $i_c = \Phi_0/[2(1 + \gamma)L_0]$ (and for large $i_c$) the periodic sign-changing variations in the current $i$ as a function of the flux $\Phi_x$ become discrete. In real low-inductance rings with point contacts, $\gamma \ll 1$. The discrete shift in the current $i$ takes place when $\Phi_x = 1/2\Phi_0$. Then the maximum negative and positive values of the current, $\delta i$, are equal to

$$\delta i = \Phi_0/2L_0, \qquad (4)$$

if the condition

$$i_c \approx \delta i = \Phi_0/(2L_0), \qquad (5)$$

is satisfied.

Equation (3) implies that in the Josephson approximation, analogous changes start at

$$i_c = \Phi_0/(2\pi L_0). \qquad (6)$$

As Eqs. (5) and (6) show, the equations are similar in both approximations. Discrete changes in the current $i$ correspond to discrete (discontinuous) changes in the magnetic flux in the circuit. It has been assumed[2] that when these conditions are satisfied the time $\delta t$ for discrete switching of the current state of a ring with a contact is comparable to the minimum possible time $\tau$ for switching of the current through the contact owing to reaching the depairing current in the contact, and the voltage $V_i$ in the circuit at that time corresponds to the energy gap of the superconducting gap $\Delta = eV_i$ eV:

$$V_i = \frac{L_0 \delta i}{\delta t} = \frac{L_0 \delta i}{\tau} = \frac{L_0 i_c}{\tau} \approx \frac{L_0 i_c}{L_0/R_N} \approx i_c R_N \approx \frac{\Delta}{e}, \qquad (7)$$

where $R_N$ is the normal resistance of the contact. In particular, using Eq. (5) we obtain

$$V_i \approx \Delta/e \approx \Phi_0/\tau. \qquad (8)$$

In the literature known to us there are no data on $V_i$ for cases where

$$i_c > \Phi_0/(2L_0) \quad \text{or} \quad i_c > \Phi_0/(2\pi L_0). \qquad (9)$$

It should be noted that the small dimensions of these superconducting rings and the small expected values of $\tau$ (e.g., for niobium $\tau \approx 10^{-12}$ s at $T = 4.2$ K) and $V_i$ have thus far prevented direct measurements of $i(\Phi_x)$, $\tau$, and $V_i$.

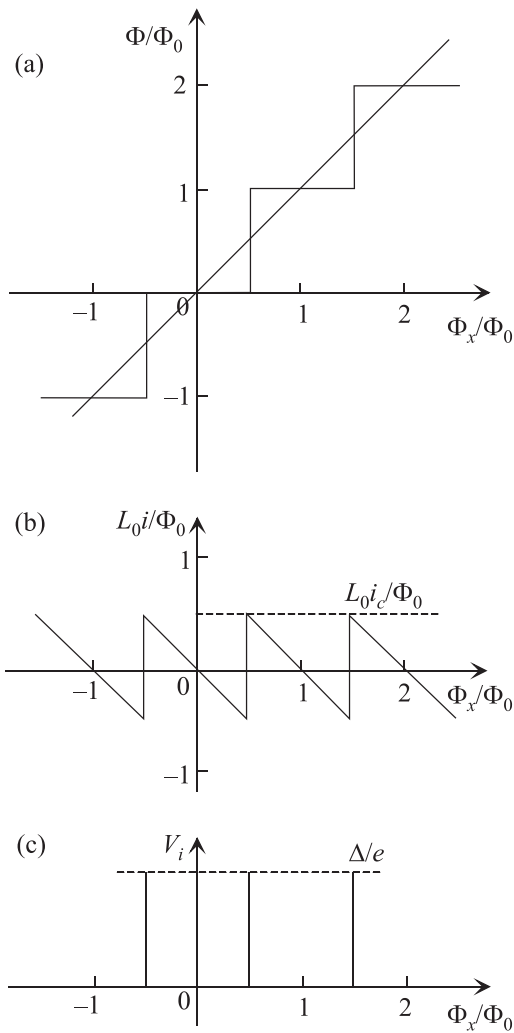

FIG. 1. Calculated dependences of the relative magnetic flux $\Phi/\Phi_0$ (a), circulating current $L_0 i/\Phi_0$ (b), and pulsed voltage $V_i$ (c) as functions of the external magnetic flux $\Phi_x$ through a low-inductance circuit with a point contact in the linear approximation of Silver and Zimmerman[2] for the case of $i_c \approx \Phi_0/(2L_0)$.

For clarity, Fig. 1 shows some dependences of $\Phi(\Phi_x), i(\Phi_x), V_i(t)$ for these processes calculated in the linear approximation.[2]

A magnetic flux $\Phi_x$ through a circuit with a contact can be created by a constant transport current $I$ through the circuit, as well as by an external magnetic field. To do this the leads for $I$ must be positioned near the contact. In this case, the part of the current $I$ flowing along the larger branch of the circuit will create $\Phi_x \neq 0$. As a result, interference occurs between the current $I$ and the circulating current $i$. Curves similar to those in Fig. 1 apply to low-inductance circuits with two parallel superconducting contacts.[4–6] In these circuits the current $I$ can produce $\Phi_x$, both because of the asymmetric inductances of its branches and because of a difference in the critical currents of the contacts. These dependences for circuits with one or two contacts have the following common properties: periodicity owing to the effect of a periodic magnetic flux $\Phi_0$ and the decisive role of the magnitude of the modulation $\delta i$ of the circulating current $i(\Phi_x)$.

Over the last few years we have studied discrete changes in the current $I_1$ in the high-inductance ($L \gg 10^{-9}$ H) branch of doubly-connected superconductor (DCS) with one and two clamped point contacts (CPC)[7–9] when a transport current $I$ flows through the DCS. Here the short branch of the DCS is a clamped contact, through which a current $I_2$ ($I = I_1 + I_2$) flows. The microstructure of the clamped point contact plays an important role in understanding the reason for these changes. Zimmerman and Silver[1] were the first to point out that a CPC is usually a unique miniature superconducting quantum interferometer (SQI) with several microcontacts with different critical points connected in parallel. In papers[7–9] dealing with a single fixed temperature (4.2 K), it is proposed that periodic discrete variations in the current $I_1$ in a DCS are a consequence of modulation of the critical current ($I_{c2}$) of the CPC-SQI by the magnetic field of the variable transport current $I$ through the DCS. This modulation is caused by quantum interference of the current $I$ through the DCS and the current $i$ circulating in the CPC-SQI. The distinctive feature of the CPC-SQI and a necessary condition for this modulation is its asymmetry, which ensure that the relation $\Phi_x = \Phi = \Phi_1 - \Phi_2 \geq \Phi_0/2$ holds between the fluxes $\Phi_1$ and $\Phi_2$ created by the currents $I_A$ and $I_B$ in the branches of the interferometer with inductances $L_A$ and $L_B$ ($L_A + L_B = L_0$) and the quantum of flux $\Phi_0$.

As opposed to low-inductance SQI, in the high-induction DCS with CPC that we have studied it is possible to make direct measurements of the amplitude of the discrete changes in the current $I_1$ (equal to the discrete changes in the circulating current $i$), their duration $\delta t$, and the corresponding voltages. This may, in turn, be a new alternative method for measuring $\Delta$ and $\tau$.

This paper is an experimental test of this possibility and clarifies the temperature dependence of the observed discrete variations in the current in high-inductance DCS with CPC.

## 2. Experimental setup

The experimental measurements were made using the DCS shown in Fig. 2(a). Figure 2(b) shows the bias circuit for this DCS in the case where the CPC consists of an SQI with two microcontacts (A and B) with different critical currents ($I_{cA}$ and $I_{cB}$). The normal resistance ($R_N$) of a contact isolated from the DCS was measured by a traditional four probe method at $T = 10$ K and found to be about $0.25$ $\Omega$.

The high-inductance DCS loop of niobium microwire contains two superconducting coils with inductances $L_3 \approx 0.6 \times 10^{-6}$ H and $L_4 \approx 10^{-6}$ H. The sum ($L$) of these inductances mainly determines the geometrical inductance of the DCS loop. The coil with inductance $L_3$ is wound on the core of a ferroprobe as described in Ref. 7. The 10-turn coil with inductance $L_4$ is the primary winding of a cryogenic pulse transformer. The transformer consists of a toroidal magnetic circuit with two identical windings. The transformer is intended to convert a jump in the current $I_1$ in the large branch of the DCS into a voltage pulse on its secondary winding. The transformer ratio is 0.5. The magnetic circuit is made of high-frequency ferrite. Ferrite was chosen after an analysis of published data on the high-frequency and temperature properties of magnetic materials[10,11] and measurements of the temperature dependence of the magnetic permeability of several brands of domestically manufactured ferrites over 300–4.2 K. The chosen ferrite has the smallest drop in permeability when the temperature is reduced to 4.2 K. The voltage pulse on the secondary winding of the transformer was recorded by an S1–83 oscilloscope with a sensitivity of 0.1 mV for a 1 mm deflection of the beam on the screen. A niobium-niobium CPC was formed at the intersection of the niobium microwires by a method described in Ref. 7. Basic information on the transport and dynamic properties of the DCS was obtained by analyzing the dependences of the current $I_1$ in the large branch of the DCS on the transport current $I$ at temperatures of 4.2–10 K and the time ($t$) dependences of the voltage pulse $V_{i4}{}^*$ on the secondary winding of the transformer at $T = 4.2$ K. Intermediate temperatures were obtained by placing the DCS in the gaseous medium of the cryostat at different distances from the liquid helium surface. In the measurements of $I_1(I)$ a constant transport current was passed through the DCS and the value of $I_1$, proportional to the magnitude of the magnetic field measured by the ferroprobe, was recorded. The $I_1(I)$ curve was recorded with a type N-309 chart recorder. $V_i(t)$ was measured by passing a low-frequency alternating transport current $I$ of varying amplitudes in excess of the critical current of the CPC through the DCS. The principle of the measurements of $V_{i4}{}^*(t)$ curves is illustrated in Fig. 4.

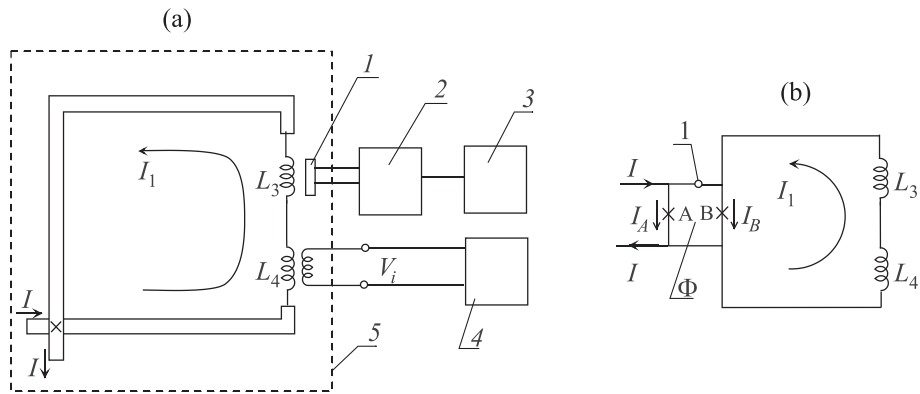

FIG. 2. (a) Measurement circuit for studying the current states of a DCS with a clamped point contact at the intersection (x) of the DCS microwires: (1) ferroprobe detector for measuring the magnetic field of the current $I_1$ through the inductor $L_3$, (2) detector signal amplifier, (3) recorder for the magnitude of the measured magnetic field, (4) oscilloscope for measuring the voltage pulses $V_i$ on the secondary winding of the transformer at the time of a discrete change in the current $I_1$ through its primary winding with inductance $L_4$, (5) cryostat. (b) Electrical bias circuit for the DCS loop in which the CPC is shown as an asymmetric superconducting quantum interferometer (1) with two microcontacts A and B with different critical currents and inserted in a branch with a different inductance; $\Phi$ is the magnetic flux through the loop of the superconducting quantum interferometer.

### 3. Experimental results and discussion

First we present the results of the measurements of $I_1(t)$ for different temperatures. Figure 3 shows plots of $I_1(t)$ for temperatures of 4.2, 8.0, and 8.5 K.

In Fig. 3(a) the $I_1(t)$ curve is similar to the dependences with periodic steps in the current $I_1$ that we observed previously at $T = 4.2$ K.[7,8] The first jump in $I_1$ takes place when the transport current $I$ reaches the critical current of the CPC ($I_{c2}$). The current jumps take place so rapidly that the recorder pen carriage overshoots to values of $I_1$ exceeding the actual step height. Then the pen returns spontaneously to the true value of the step (one such segment is indicated in Fig. 3(a) by a circle). $I_{c2}$ becomes smaller for $T > 4.2$ K in the $I_1(I)$ curves and the shape of the steps changes and they finally disappear (Fig. 3(c)).

Before explaining the effect of temperature on the $I_1(t)$ curves, let us examine our model for the formation of this curve. We know from our past experiments that the height ($\delta I_1$) and width ($\delta I$) of the steps in the $I_1(t)$ curves are equal to one another. Given the periodic character of these dependences, we can assume that each of these currents corresponds to the formation of a quantum of magnetic flux $\Phi_0$ in an interferometer containing two microcontacts. This means that the current $\delta I$ creates the exciting magnetic flux and the current $\delta I_1$ is equal to the maximum value of the current $\delta i$ circulating in the interferometer and is the diamagnetic response to the exciting flux. The flux of this response is equal to the magnitude of the exciting flux.

These fluxes are created by currents flowing along identical segments of the niobium microwires that form the CPC. This ensures that $\delta I = \delta I_1$. It can be seen in Fig. 3(a) that $I_{c2}/\delta I_1 \approx 3$. The transport current $I$ serves two functions: it ensures that the state of the CPC is close to critical and creates a magnetic flux $\Phi = \Phi_x$ that acts on the CPC-SQI circuit. In the superconducting state of the CPC the current $I$ is distributed over the two branches of the DCS in inverse proportion to their inductances $L_1$ and $L_2$. Since $L_1 \gg L_2$, the current $I$ flows mainly through the CPC. When a current $I$ close to $I_{c2}$ is reached, according to Eq. (4) the circulating current $i$ reaches its maximum in a time comparable to $\tau$. Over this time resistance develops in the CPC. This causes switching of a part $\delta I = \delta I_1$ of the transport current into the large superconducting branch of the DCS by an amount

$$\delta I = \delta I_1 = \delta i = \Phi_0/2L_0. \qquad (10)$$

The current $I$ through the CPC-SQI decreases by $\delta I$ and the contact again becomes superconducting. With further increase in the current $I$ and the corresponding flux $\Phi = \Phi_x$, the circulating current decreases as shown in Fig. 1(b). Here the current $I_1 = \delta I_1$ cannot change because of the conservation of magnetic flux in the large loop of the DCS and remains unchanged until there is a new jump in the interferometer current $i$ as $\Phi$ increases. Thus, after $\Phi$ changes by one quantum of flux the picture is repeated and the next step in the current $I_1$ appears. It is clear from this description of the changes in the current state of the DCS that the transport current $I_2$ through the CPC never reaches the critical value. It only keeps the state of the contact close to critical for all values of $I$ at which discrete changes in $I_1$ are observed in the $I_1(I)$ curve. The closeness to the critical state constitutes a fraction $\delta i$ (10). This fraction cannot be determined using the classical model for the distribution of the current $I$ in a circuit with parallel branches, since discrete switching of part of it into the branch with inductance $L_1$ is a purely quantum mechanical effect.

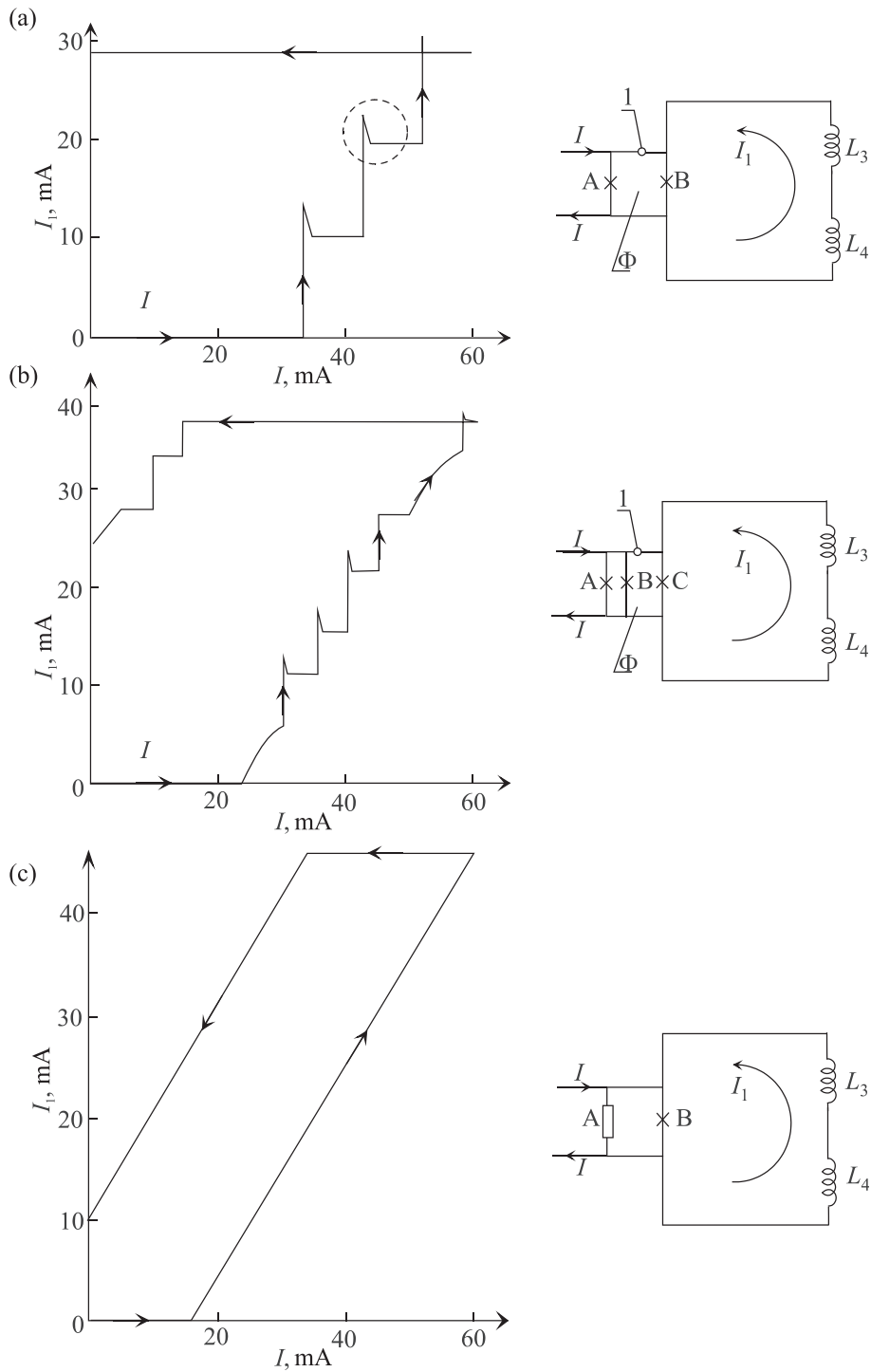

FIG. 3. The current $I_1$ in the large branch of the DCS as a function of the transport current $I$ through the DCS at temperatures of 4.2 (a), 8.0 (b), and 8.5 K (c). The bias circuits of the DCS is shown on the right. In these circuits 1 denotes the interferometer circuit, $\Phi$ is the magnetic flux acting on the interferometer, and A, B, and C are the microcontacts in the interferometer structures for the circuits of (a) and (b); in the circuit (c) A denotes the resistance of the microcontact A.

The critical state with the appearance of a resistive contact is, therefore, only caused by a quantum jump by $\delta i = i_c$ in the current $i$.

Equation (10) can be used to estimate $L_0$. After substituting the experimental values $\delta I = \delta I_1 = 9$ mA, we obtain $L_0 \approx 10^{-13}$ H. To test the agreement between the inductance of our CPC-SQI and the calculated value, we estimate it from the dimensions of the SQI circuit. The diameter $d$ of a clamped point microcontact can be estimated using the expression[12]

$$d \approx \rho/R, \qquad (11)$$

where $\rho$ is the resistivity of niobium. For $\rho = 10^{-8}$ $\Omega \cdot$m and $R_N \approx 0.25$ $\Omega$, we obtain $d \approx 4 \times 10^{-8}$ m. Since the cross section s of the microcontact and its length $t$ are related by $s \approx d^2 \approx t^2$ (Ref. 2) and the distance between the microcontacts of the SQI can be comparable to its diameter, the effective diameter of the SQI loop is comparable to the diameter $d$. The approximate inductance $L_0$ of an SQI loop with diameter $d$ is given by[13]

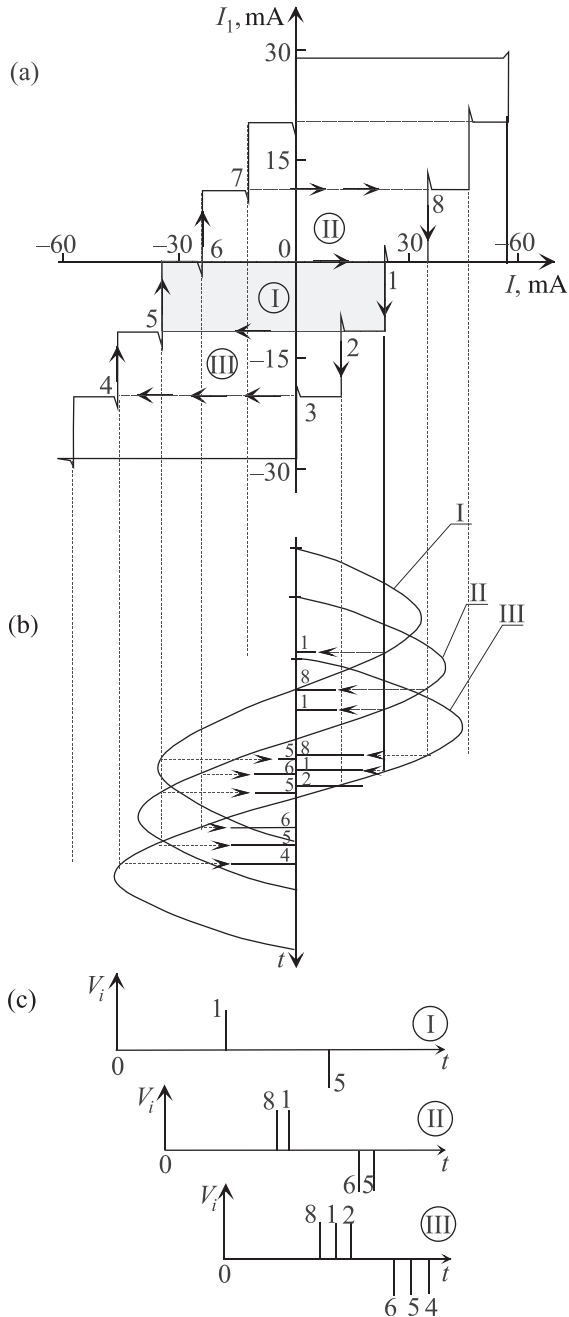

FIG. 4. Scheme for obtaining different numbers of voltage pulses $V_{i4}^*$ on the oscilloscope screen with an alternating transport current $I$ through the DCS: (a) experimental dependence $I_1(I)$ at $T = 4.2$ K; (b) variation in the current $I$ during a single period for three different amplitudes I, II, III; (c) the expected pattern of the voltage pulses on the oscilloscope screen (1–8 denote the numbers of the steps in the current $I_1$ excited for different amplitudes of the current $I$).

$$L_0 \approx 4\pi \times 10^{-7} d. \tag{12}$$

Equation (12) implies that $L_0 \approx 4 \times 10^{-14}$ H. This is close to $10^{-13}$ H, which means that conditions (4) and (10) are satisfied approximately; these conditions are necessary for the existence of discrete transitions in the circulating current in our SQI-CPC and the corresponding step dependences of $I_1(I)$ with changes in the transport current through the DCS.

The above results can be used to test the proposed model for the periodic step dependence of $I_1(I)$ in Fig. 3(a). The complete absence of steps in Fig. 3(c) at $T = 8.5$ K can be explained by the disappearance of superconductivity in the weaker of the two microcontacts, which form an interferometer at $T = 4.2$ K and not in the one with a higher critical current. This leads to termination of the periodic modulation of the critical current of the CPC by the magnetic field of the current $I$. The critical current of the DCS becomes equal to the critical current ($I_{c2} \approx 16$ mA) of the only superconducting contact of the CPC. The shape of the $I_1(I)$ current for $T = 8.0$ K (Fig. 3(b)) is intermediate between the periodic current steps at $T = 4.2$ K and the curve without any steps at $T = 8.5$ K. When $I > I_{c2}$ this curve contains segments with and without steps. A possible explanation for this nonuniform dependence may be the formation of CPC-SQI with more than two different microcontacts at a given temperature. Further study is needed for a conclusive determination of the reasons for this kind of dependence.

The transformation of the $I_1(I)$ curves at different temperatures is, therefore, a confirmation of the decisive role of the properties of the CPC as a quantum interferometer in the appearance of discrete current states in a high-inductance DCS. This transformation also confirms the assumption that there are different critical currents for the constituent microcontacts in the interferometer. In sum, we can say that it is possible to choose a CPC with the structure of a two-contact SQI by varying the temperature.

We now turn to an experimental study of the parameters of the voltage pulse during discrete changes in the current $I_1$ in the $I_1(I)$ curve at $T = 4.2$ K. Figure 4 illustrates the method for obtaining voltage pulses $V_{i4}^*$ on the oscilloscope screen when alternating transport currents $I$ with different amplitudes (I, II, III) are passed through the DCS. The frequency of the alternating current is 100 Hz.

Figure 4 shows that, for the lowest amplitude (I) of the current, discrete changes in $I_1$ occur in the form of a sequence of steps 1 and 5 in the current $I$. For amplitude II the changes in $I_1$ occur with successive flow of the current $I$ in steps 8, 1, 5, and 6. For amplitude III the changes in $I_1$ occur with successive passage of steps 8, 1, 2, 6, 5, and 4. During each change in $I_1$, voltage pulses

$$V_{i3} = L_3 \delta I_1/\delta t, \tag{13}$$

and

$$V_{i4} = L_4 \delta I_1/\delta t, \tag{14}$$

appear on the inductances $L_3$ and $L_4$, where $\delta t$ is the pulse duration. The voltage $V_{i4}$ is converted into a voltage $V_{i4}^*$ on the secondary winding of the transformer which is recorded on the oscilloscope screen.

Figure 5 shows some oscilloscope traces of the voltage pulses on the secondary winding of the cryogenic transformer for different amplitudes of the low-frequency transport current through the DCS at $T = 4.2$ K.

A comparison of Figs. 4 and 5 demonstrates the consistency of the computational and experimental pictures of the origin of the pulsed voltage at the times of the jumps in the currents $i$ and $I_1$.

Rescaling the detected amplitude of the pulsed voltage $V_{i4}^*$ to the voltage $V_i^{DCS} = V_{i3} + V_{i4}$ on the entire DCS circuit, including the transformer ratio for this voltage into the voltage on the primary winding $V_{i4}$ and Eqs. (13) and (14), gives $V_i^{DCS}(4.2\,\text{K}) = (2.6 \pm 0.2)$ mV. The experimental value is close to the gap for niobium measured by a tunnel method (2.8 meV).[14] The temperature dependence $V_i^{DCS}(T)$ is shown in Fig. 6 as a plot of the experimental data and the known dependences of the relative gap $\Delta(T)/\Delta(0)$ as functions of relative temperature $T/T_c$.[14] The experimental values are close to the niobium gaps measured by tunnelling. The duration ($\delta t$) of the voltage pulse is in the range of $10^{-6}$–$10^{-5}$ s. The accuracy with which the shape and duration of the pulses are recorded on the oscilloscope is limited by the level of noise reduction in the measurement circuit. The average duration of the pulse can be found using the formula

$$V_i^{DCS} \approx L \delta I_1 / \delta t. \quad (15)$$

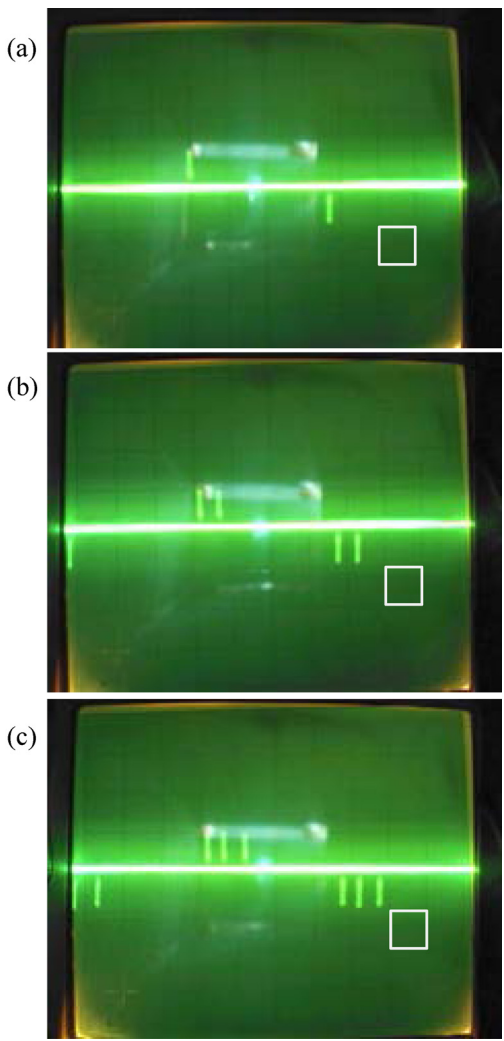

FIG. 5. Oscilloscope traces of the voltage pulses $V_{i4}*$ on the secondary winding of the transformer inserted in the DCS circuit for three values of the amplitude of the 100 Hz alternating current $I$ ($I_a < I_b < I_c$). (The horizontal edges of the white squares correspond to $10^{-3}$ s and the vertical edges, to 1 mV.)

Substituting the experimental values of $V_i^{DCS}, L, \delta I_1$ gives $\delta t \approx 0.6 \times 10^{-5}$ s.

It should be noted that a calculation of the time constant $(L/R_N)$ of the DCS circuit at the time of a jump in the current $I_1$ and of the resistance of the CPC gives a value consistent with Eq. (15).

These results yield the following conclusions. The amplitude of the voltage pulses on the DCS and CPC is close to the energy gap for niobium.[14] In this regard, the measured value ($V_i^{DCS}$) is close to the measured gap singularity in the current-voltage characteristic of an Nb-I-Nb tunnel contact.[14] At the same time, we have the question of why does $V_i^{DCS}$ differ so much from the traditional estimate of the gap given by $\Delta/e \approx I_{c2} R_N$? If we replace $I = I_{c2} \approx 30$ mA by $\delta I_1/2 = i_c \approx 4.5$ mA from our experiment, we get $\Delta/e = 1.15$ mV, which is close to $\Delta/e$ for niobium[14] at $T = 4.2\,\text{K}$ ($\Delta/e = 1.4$ mV). The correctness of this substitution is explained by the fact that in our experiments the transport current $I$ serves two functions: it creates a magnetic flux $\Phi = \Phi_x$ through the CPC-SQI and brings the SQI to a critical state that differs from it by $\delta i \approx i_c$. This can be represented using Fig. 1(b) if the origin for $\Phi_x/\Phi_0$ is taken at $I \approx I_{c2} - i_c$. As $\Phi_x/\Phi_0$ increases from 0 to 1/2 the absolute magnitude of $i$ rises to the critical state of the SQI with a discrete change in $i$ by $2i_c$. This shows that a circulating current of Cooper pairs $i$, in other words a periodic process of electron depairing followed by pairing at $\Phi_x/\Phi_0 = (k + 1/2)$, where $k = 0, 1, 2, 3,…$, controls the periodic change in the critical state of the CPC-SQI, and the transport current in this case is an auxiliary participant in the process.

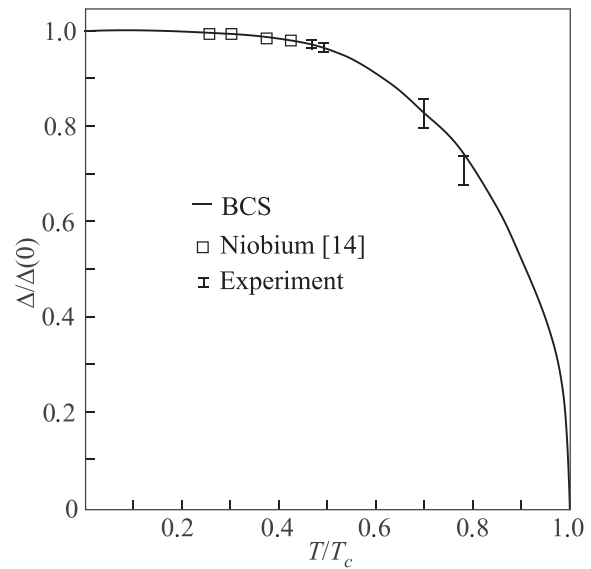

FIG. 6. Relative values of the energy gap of niobium as a function of the relative temperature according to the BCS theory and experimental values of the gap derived from measurements of the characteristics of tunnel contacts[14] (☐) and from our measurements of voltage pulses $V_i^{DCS}$ on DCS.

Therefore, the proposed technique of simultaneously measuring the amplitude and duration of the voltage pulse on the DCS may be an new alternative way of measuring the energy gap of superconductors and the time for changes in the discrete current state of quantum interferometers.

The duration of the voltage pulse on the DCS (15) is many orders of magnitude greater than the value calculated for an isolated quantum interferometer[2] ($\tau \sim 10^{-12}$ s). This differences is caused by the much larger inductance of the DCS circuit ($L \sim 10^{-6}$ H) compared to the inductance of an isolated interferometer ($L_0 \sim 10^{-13}$ H). In fact, given the equality of $V_i^{DCS}$ (15) and $V_i$ (7) in the structure of a DCS, it follows that $L\delta I_1/\delta t = L_o i_c/\tau$. Since $\delta I_1 \approx i_c$, we find $\delta t/\tau = L/L_0$, which means that

$$\tau \approx (L_0/L)\delta t. \quad (16)$$

This value of $\delta t$ is in the radio-frequency range of the observed voltage pulses on a DCS. It follows from Eq. (16) that

$$\tau_{Nb} \approx 10^{-12} c, \quad (17)$$

which is close to a calculated estimate of this quantity.[2]

It also follows from the experiment that the power $P$ of each voltage pulse on the DCS is equal to $P \approx V_i^{DCS} \delta I_1 \approx 2 \times 10^{-5}$ W. Equation (16) implies that the pulse duration ($\delta t$) of the voltage on the DCS can be controlled by changing the ratio of the inductances of the DCS and the interferometer.

The accuracy of a simultaneous determination of the energy $\Delta$ of a discrete state of a quantum system and its lifetime $\tau$ is limited by the Heisenberg uncertainty relation $\Delta \cdot \tau \approx h$. The experimental data obtained here confirm this. Substituting $\Delta_{Nb}(4.2\,K) \approx (1.4 \times 10^{-3}\,V)(1.6 \times 10^{-19}\,C) = 2.2 \times 10^{-22}$ J and $\tau \approx 10^{-12}$ s yields $\sim 2.2 \times 10^{-34}$ J·s, a value close to $h$.

We have, therefore, shown for the first time that the transformation of quantum discrete values of the current in a superconducting interferometer into discrete values of the current in the large shunt inductance of a DCS loop can be used for simultaneous measurement of the superconductor gap and the duration of the voltage pulses on the DCS. The measured pulse durations can be used to estimate the minimum possible time for a quantum discrete transition in the CPC of the interferometer. The duration of the transition determines the maximum attainable speed of superconducting computing devices.

Successful measurement of these parameters became possible because of discrete processes in a combined high-inductance DCS+ quantum interferometer structure at lower frequencies than in an isolated interferometer. This, in turn, has made it possible to use simpler and more accessible low-frequency means of measurement.

It should be noted that discrete changes in the current state are also observed in DCS with CPC during changes in an external magnetic field.[9] This makes it possible to measure the energy gap of superconductors using a DCS without activating it from a source of constant transport current.

The data obtained here on the parameters of the voltage pulses $V_i^{DCS}$ are an important addition to our earlier[7] model of discrete changes in the current states of DCS with CPC and confirm our assumption about switching of the transport current into the high-inductance loop of a DCS in connection with a discrete change in the circulating current in a CPC-SQI acted on by a magnetic flux created by the transport current through the DCS.

### 4. Conclusions

In an experimental study of a high-inductance doubly connected superconductor with a niobium-niobium clamped point contact we have found that temperature has a significant effect on the form of the dependences of the current $I_1$ through the large branch of a DCS on the transport current $I$ through the DCS. This effect is explained by a difference in the temperature dependences of the critical currents of the microcontacts in the CPC in the 4.2–9 K range. In particular, at $T = 4.2$ K a clamped point contact contains only two parallel superconducting microcontacts that form a superconducting quantum interferometer with the necessary parameters for studying discrete current changes.

The conversion of discrete changes in the current in the large branch of a DCS into a pulsed voltage on that branch using a cryogenic transformer has shown that it is close to the energy gap of niobium. This new method of obtaining information on the energy gap of superconductors and its temperature dependence may be an alternative (and in some cases, simpler) to the earlier method (in particular, the tunnelling and infrared absorption methods). At the same time, we have determined the range of the duration of the voltage pulse on a DCS during discrete change in the quantum state of a CPC-SQI. The measured durations $\delta t$ of the change in the discrete change in the current state of a DCS with a CPC-SQI turned out to be many orders of magnitude greater than the expected $\tau$ for an isolated quantum interferometer. This means that the switching of the transport current from a quantum interferometer into the high-inductance branch of a DCS is considerably shorter than the assumed switching rate during a quantum change in the current state of an isolated low-inductance interferometer. It has been shown that the slowing down factor is proportional to the ratio of the inductances of the DCS loop and the interferometer itself. Scaling of $\delta t$ to $\tau$ with this circumstance taken into account has made it possible to obtain a previously predicted, but unmeasured, value of $\tau$ ($\sim 10^{-12}$ s). Further improvement in the technique for recording the observed voltage pulse should yield more precise values of these parameters.

We thank A. N. Omed'yanchuk for useful discussions of the results in this paper, N. I. Pyatak for interest in this work and help in choosing the high-frequency ferrites, and N. G. Burma and E. V. Kovaleva for help in making the ferrite-based pulse transformer.

a)Email: bondarenko@ilt.kharkov.ua